\documentclass[twocolumn,superscriptaddress,showpacs,floatfix,amsmath,amssymb]{revtex4-1}
\usepackage{graphicx}
\usepackage[dvipdfmx,colorlinks=true,linkcolor=blue,citecolor=blue,urlcolor=blue]{hyperref}
\usepackage[utf8]{inputenc}
\usepackage{float}

\newcommand{\singlet}{{}^1\mathrm{S}_0}
\newcommand{\triplet}{{}^3\mathrm{P}_2}
\newcommand{\Li}{{}^2\mathrm{S}_{1/2}}

\newcommand{\cm}{\mathrm{cm}}
\newcommand{\micron}{\mu\mathrm{m}}
\newcommand{\nm}{\mathrm{nm}}
\newcommand{\msec}{\mathrm{ms}}
\newcommand{\s}{\mathrm{s}}
\newcommand{\G}{\mathrm{G}}

\newcommand{\kHz}{\mathrm{kHz}}
\newcommand{\MHz}{\mathrm{MHz}}

\begin{document}

\title{Spectroscopic determination of magnetic-field-dependent interactions in an ultracold Yb($\triplet$)-Li mixture}

\author{F.~Sch\"{a}fer}
\email{schaefer@scphys.kyoto-u.ac.jp}
\affiliation{Department of Physics, Graduate School of Science, Kyoto University, Kyoto 606-8502, Japan}

\author{H.~Konishi}
\affiliation{Department of Physics, Graduate School of Science, Kyoto University, Kyoto 606-8502, Japan}

\author{A.~Bouscal}
\affiliation{Department of Physics, Graduate School of Science, Kyoto University, Kyoto 606-8502, Japan}
\affiliation{Department of Physics, \'{E}cole Normale Sup\'{e}rieure, 24 Rue Lhomond, F-75231 Paris Cedex 05, France}

\author{T.~Yagami}
\affiliation{Department of Physics, Graduate School of Science, Kyoto University, Kyoto 606-8502, Japan}

\author{Y.~Takahashi}
\affiliation{Department of Physics, Graduate School of Science, Kyoto University, Kyoto 606-8502, Japan}

\date{\today}

\begin{abstract}
	We present experimental results on the inelastic and elastic interspecies
	interactions between ytterbium (Yb) in the metastable $\triplet$ state
	loaded into a deep optical lattice and spin polarized lithium (Li) in its
	ground state. Focusing on the $m_J = 0$ magnetic sublevel of Yb($\triplet$),
	bias magnetic fields between $20~\G$ and $800~\G$ are investigated and
	significantly enhanced inelastic collision rates with high magnetic fields
	are found. In addition, by direct spectroscopy of the Yb Mott-insulator immersed
	in the Li Fermi gas an upper boundary of the background scattering length of
	the Yb($\triplet, m_J=0$)-Li($\Li, F=1/2, m_F=+1/2$) system is estimated,
	revealing the absence of useful Feshbach resonances. These observations are
	qualitatively consistent with the theoretical calculations.
\end{abstract}

\maketitle

\section{Introduction}

Experiments in ultracold quantum degenerate gases have proven to be an
invaluable asset in the study of quantum phenomena~\cite{bloch_many-body_2008,
bloch_quantum_2012}. Especially the defect-free preparation of samples in
optical lattices~\cite{giorgini_theory_2008} allowed for unprecedented control
in realizing the Bose-Hubbard model~\cite{bloch_ultracold_2005} and is seen as
a promising platform to realize topological
matter~\cite{goldman_topological_2016}. Success is facilitated by an
unparalleled controllability of the particle interactions by means of
magnetically tunable Feshbach resonances~\cite{chin_feshbach_2010}. In recent
years multi-component quantum gases also came into
focus~\cite{myatt_production_1997} with applications in quantum simulation of
impurity systems~\cite{massignan_polarons_2014} and in the production of
ultracold molecules~\cite{carr_cold_2009, moses_new_2017}. Large progress was
made in the formation of ground state alkali-metal
dimers~\cite{jones_ultracold_2006}, and their applications to ultracold
chemistry~\cite{ospelkaus_quantum-state_2010} and dipolar collisional
physics~\cite{ni_dipolar_2010} were reported.

Shifting away from alkali-metal dimers to compounds including alkaline-earth
or alkaline-earth-metal-like species a new class of molecules with doublet
ground states ($^2\Sigma$ molecules) becomes possible. New applications in
quantum state preparation~\cite{perez-rios_external_2010} and information
processing~\cite{micheli_toolbox_2006} are envisioned. Production approaches
include the photoassociation of molecules~\cite{jones_ultracold_2006,
nemitz_production_2009}, buffer gas loading
techniques~\cite{weinstein_magnetic_1998, maussang_zeeman_2005} and direct
laser cooling~\cite{shuman_radiative_2009}. Mandatory to the formation of
ultracold $^2\Sigma$ molecules is a good understanding of the interspecies
interactions and the control thereof~\cite{krems_cold_2009}. The possibility
of magnetically tunable Feshbach resonances in those systems was confirmed
theoretically~\cite{zuchowski_ultracold_2010}, however, the predicted
resonances are very narrow~\cite{brue_magnetically_2012}. Later it was
discussed that due to anisotropy in the interaction involving the metastable
$\triplet$ state~\cite{reid_fine-structure_1969} possibly experimentally
exploitable, broad magnetic Feshbach resonances are supported by the ytterbium
(Yb)-lithium (Li) collisional system, where the alkaline-earth-metal-like Yb
is in the metastable $\triplet$
state~\cite{gonzalez-martinez_magnetically_2013, petrov_magnetic_2015,
chen_anisotropy_2015}. At the same time, those theories predict quite large
inelastic two-body collisional processes and thus the suppression of a large
variation of the scattering lengths around the Feshbach resonances.

Due to the complexity of the calculations involved and insufficient knowledge
on the precise interaction potentials, reliable predictions on the
Yb($\triplet$)-Li Feshbach resonance landscape are as of yet missing.
Experimentally, a first signature of a Feshbach resonance in $^{174}{\rm
Yb}(\triplet, m_J = -1)$-$^6{\rm Li}(\Li, F=1/2, m_F=+1/2)$ collisions was
found~\cite{dowd_magnetic_2015} leading to improved model
calculations~\cite{petrov_magnetic_2015}. In a different
work~\cite{konishi_collisional_2016} the inelastic loss coefficient in the
energetically lowest state, $^{174}{\rm Yb}(\triplet, m_J = -2)$-$^6{\rm
Li}(\Li, F=1/2, m_F=+1/2)$, at low magnetic bias fields and the involved
inelastic loss channels were discussed. Most recently, the importance of
anisotropy induced losses and their possible suppression using stretched
states was experimentally highlighted~\cite{schafer_spin_2017}.

The purpose of the present work is to further broaden our knowledge on the
$^{174}{\rm Yb}(\triplet)$-$^6{\rm Li}(\Li)$ collisional system. We
experimentally investigate the inelastic losses in collisions of $^{174}{\rm
Yb}(\triplet, m_J = 0)$ with ground state $^6{\rm Li}$ in the Zeeman states
$F=1/2,\ m_F = \pm 1/2$ and $F=3/2,\ m_F = \pm 3/2$ for bias magnetic fields
between $20~\G$ and $800~\G$. Further, by direct spectroscopic investigation
of the ultranarrow Yb($\singlet \rightarrow \triplet$) transition we gain
direct access to the elastic part of the interspecies scattering length and
estimate an upper bound for it.

This paper is organized as follows. In Sec.~\ref{sec:expt}, we review the
experimental idea and details on its execution. Section~\ref{sec:results}
presents the data and its analysis. In Sec.~\ref{sec:discussion} we conclude
by a discussion of the results and their significance.

\section{Experiment}
\label{sec:expt}

The experiment started with a degenerate mixture of bosonic $^{174}$Yb and
fermionic $^6$Li as detailed in~\cite{hara_quantum_2011, schafer_spin_2017}.
(In the remainder of the present text we will suppress the atomic mass
notation for clarity.) In brief, by combination of optical pumping of Li,
forced evaporation of Yb in a far off-resonance optical trap (FORT) and
sympathetic cooling of Li by Yb, quantum degenerate gases of typically $10^5$
Yb atoms and $3\times10^4$ spin polarized Li atoms were created and held in a
crossed FORT configuration. Li atoms could be prepared with purities $> 90\%$
in four different Zeeman states, $F=1/2, m_F = \pm 1/2$ and $F=3/2, m_F = \pm
3/2$. The temperature of Li was $T_{\rm Li} \approx 300~{\rm nK}$ and
$T_{\rm Li}/T_{\rm F} \approx 0.2$, where $T_{\rm F}$ is the Fermi
temperature.

\begin{figure}[tb]
	\centering
	\includegraphics[width=7.5cm]{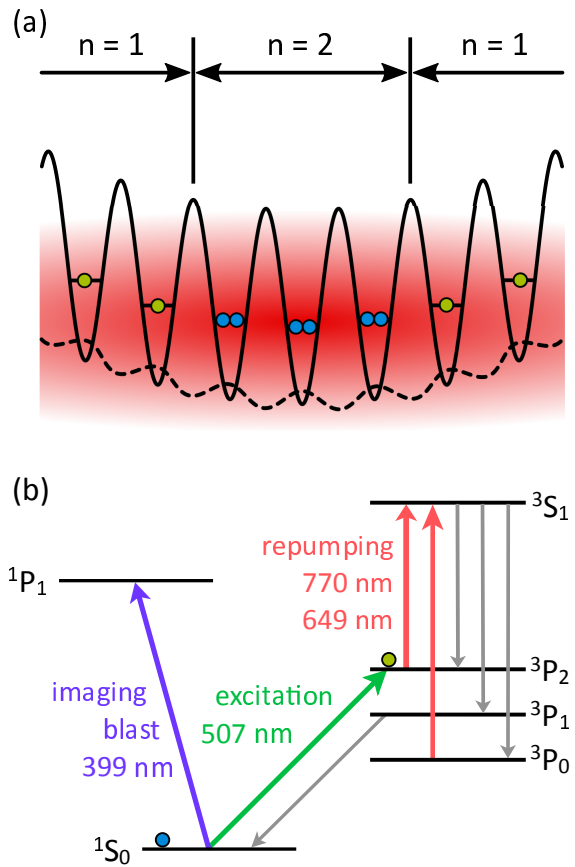}
	\caption{
		(a) Quantum degenerate mixture of Yb and Li loaded into a 3D optical
		lattice. The effective lattice potential for Yb (solid line) is deep at
		$15\,E_R^{\rm Yb}$ and a Mott-insulator shell structure (blue and green
		dots) is formed. For clarity only lattice occupation numbers $n = 1$ and
		$2$ are shown. The effective Li potential (dashed line) is with only
		$0.7\,E_R^{\rm Li}$ shallow and Li unlocalized (red cloud). In the
		experiment the interactions between Yb($\triplet$) atoms (green dots) in
		$n = 1$ sites and Li are investigated.
		(b) Yb level spectrum of importance to the experiment. The excitation
		laser (green) connects the ground to the metastable $\triplet$ state. For
		imaging purposes (blue) two repumping lasers (red) are used to repopulate
		the $\singlet$ state. 
	}
	\label{fig:setup}
\end{figure}

The experimental idea for the precise and magnetic sublevel resolved
spectroscopic determination of the Yb($\triplet$)-Li interactions is sketched
in Fig.~\ref{fig:setup}(a). The necessary Yb optical transitions are
summarized in Fig.~\ref{fig:setup}(b). Similar to our previous
work~\cite{schafer_spin_2017} the Yb atoms were prepared in a Mott-insulating
state by adiabatic loading of a three-dimensional optical lattice with lattice
constant $266~\nm$. The lattice depth with respect to Yb($\singlet$) was
$15~E_R^{\rm Yb}$, where $E_R$ denotes the recoil energy, and was for
Yb($\triplet$) due to different polarizabilities a factor $1$--$1.4$ deeper.
By virtue of its low mass the lattice depth for Li was only $0.7~E_R^{\rm Li}$
and the Li density distribution was only weakly modulated by the
lattice. At the lattice wavelength the polarizability of Li is negative and
the lattice sites of Li and Yb alternate. By application of a gravitational
sag compensation beam~\cite{konishi_collisional_2016} while ramping up the
lattice the final separation between the Yb and Li cloud center-of-mass
positions could be reduced to $(3.5 \pm 1.0 )~\micron$. Typical atom
cloud radii in vertical direction were $3.5~\micron$ for Yb and $4.0~\micron$
for Li, and the remaining center-of-mass difference was accounted for in the
data analysis.

For measurements of the interspecies inelastic decay described in detail in
Sec.~\ref{sec:inel} only Yb atoms in singly occupied lattice sites ($n = 1$)
were selectively transferred into the metastable $\triplet$ state by a
\mbox{$0.5$-$\msec$} pulse of \mbox{$507$-$\nm$} laser light. The laser
frequency was during this time linearly ramped from $-4~\kHz$ to $+4~\kHz$
with respect to the transition resonance frequency in order to improve
excitation reproducibility. By virtue of the Yb-Yb intraspecies interaction
the resonance frequencies for Yb($\singlet$) atoms in $n>1$ sites were shifted
by more than $4~\kHz$ and accordingly not excited~\cite{kato_laser_2016}. Care
was taken to choose the light intensity such as to only excite about
$2\times10^3$ to $3\times10^3$ Yb atoms, corresponding to about $10\%$ of the
total number of Li atoms. Sample preparation was completed by removal of
remaining Yb($\singlet$) atoms via a \mbox{$0.3$-$\msec$} resonant light pulse
at $399~\nm$. This ensured that due to Yb residual site hoppings a decay of
metastable Yb atoms by collisions with Yb($\singlet$), known to be highly
inelastic~\cite{uetake_spin-dependent_2012}, was excluded.

Two magnetic coils in Helmholtz configuration provide up to \mbox{$800$-$\G$}
magnetic bias field at the position of the atoms. Magnetic field calibration
was done by observation of the narrow Li Feshbach resonance at
$543~\G$~\cite{schunck_feshbach_2005} and linear extrapolation. Linearity was
guaranteed by a temperature stabilized closed-loop feedback system of the coil
current. The residual uncertainty in the magnetic fields was estimated to be
$0.5\%$ of the nominal value. It was further confirmed that possible effects
of magnetic field inhomogeneities on the position of the magnetically
sensitive Li atom cloud were smaller than the uncertainty in the final,
gravity induced Yb-Li cloud separation. To prevent detrimental effects of a
magnetic field ramp on the measurement results it was desirable to set the
magnetic field before excitation to the metastable state. We note that in fact
we found that a field ramp-up sequence after excitation to the $\triplet$
state at low magnetic field introduced severe instability of the signals,
possibly due to the metastable state already decaying during the field ramp-up
process. Due to the strong Zeeman splitting of the metastable Yb state of
$2.1~\MHz/\G \times m_J$ the constraint of an excitation only at high fields
limited the choice of possible Zeeman states to the magnetically insensitive
$m_J = 0$ one. At magnetic field strengths beyond about $300~\G$ excitation to
the $\triplet$ state was facilitated by magnetic admixture of the $^3{\rm
P}_1$ state. For low fields and due to experimental constraints in our choice
of relative field polarizations efficient excitation was not possible.
Therefore we chose to perform for target fields below \mbox{$300$-$\G$}
excitation to the metastable state at $300~\G$ followed by a fast
$0.3$-$\msec$ ramp (the duration of the \mbox{$399$-$\nm$} sample cleaning
pulse) down to the target field. In all cases the upwards ramp after loading
Yb into the lattice and before excitation to the metastable state was
$20~\msec$ in duration.

After sample and magnetic field preparation a variable holding time was
followed by a second cleaning pulse at $399~\nm$. This guaranteed that only
the remaining Yb($\triplet$) atoms were detected after the interaction time
by preventing any spontaneous or collision induced Yb($\triplet
\rightarrow \singlet$) events not leading to trap loss from contaminating
the experimental signal. Detection of the remaining atoms was done after
repumping to the $\singlet$ ground state, Fig.~\ref{fig:setup}(b), by
recording the fluorescence light of a detection MOT operating on the strong
$\singlet$-$^1{\rm P}_1$ transition at $399~\rm{nm}$. The number of Li atoms
was simultaneously recorded by standard absorption imaging. Repetition of the
complete sequence for interaction times typically between $0$ and $30~\msec$
gave access to the necessary decay information.

In a second set of measurements, see Sec.~\ref{sec:el}, the on-site
interaction energy between Li and Yb($\triplet$) was probed directly by
measuring the induced shift of the Yb($\singlet \rightarrow \triplet$)
transition energy~\cite{konishi_collisional_2016}. This required two
modifications to the experimental sequence. First, excitation to the
metastable state was done with a \mbox{$0.3$-$\msec$} pulse of constant
frequency. Second, after the first cleaning pulse no further interaction time
was necessary and the number of Yb($\triplet$) atoms was directly recorded by
fluorescence imaging. Repeating the modified sequence with excitation
frequencies typically $\pm 10~\kHz$ about the resonance frequency revealed the
$n = 1$ excitation spectrum.

In both cases the measurements were completed by taking additional reference
data where the cloud of Li atoms was removed from the experiment by
application of a strong blasting light resonant to the Li D2 line after
completing the forced evaporation and before loading the atoms into the
optical lattice. All measurements were repeated five times.

\section{Analysis and results}
\label{sec:results}

In the following we show the experimental data obtained and present their
analysis. The results concerning the inelastic collisional properties are
treated in Sec.~\ref{sec:inel}, those regarding the elastic interactions are
elucidated in Sec.~\ref{sec:el}.

\subsection{Inelastic losses}
\label{sec:inel}

\begin{figure}[tb]
	\centering
	\includegraphics[width=8cm]{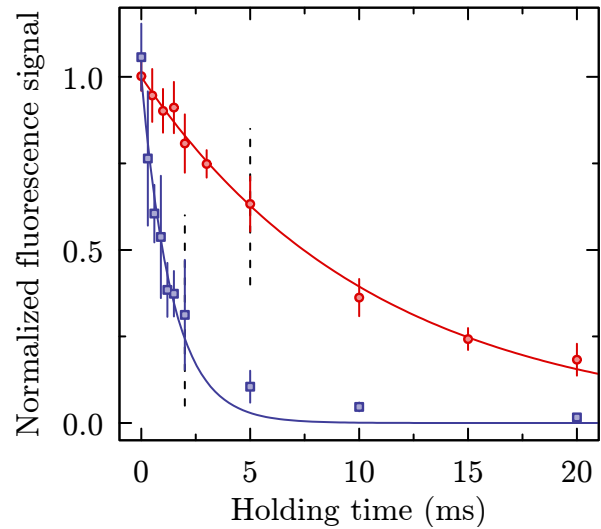}
	\caption{
		Experimental decay signal and lifetime determination. For the inelastic
		Yb($\triplet, m_J = 0$)-Li($F = 1/2, m_F = +1/2$) collisional process the
		decay of metastable Yb atoms is shown at magnetic bias field strength of
		$20~\G$ (red circles) and $600~\G$ (blue squares). Lines indicate
		exponential fits to the data where data points up to $5~\msec$ and
		$2~\msec$ (dashed lines), respectively, were evaluated. Lifetimes were
		found to be $10.7^{+1.3}_{-0.8}~\msec$ and $1.4^{+0.2}_{-0.1}~\msec$. The
		error bars indicate the standard deviation over typically five independent
		executions of the measurement.
	}
	\label{fig:decay}
\end{figure}

We measured the magnetic field and spin dependence of the inelastic loss
coefficient by observation of the Yb($\triplet$) decay curve for each of the
four accessible Li Zeeman states and magnetic bias fields between $20$ and
$800~\G$ in steps of $20~\G$. As an example, the obtained decay signals for
the Yb($\triplet, m_J = 0$)-Li($F = 1/2, m_F = +1/2$) collisional process at
$20~\G$ and $600~\G$ are reproduced in Fig.~\ref{fig:decay}. A significantly
shorter lifetime at $600~\G$ is observed. For the analysis we start with the
usual ansatz~\cite{konishi_collisional_2016, schafer_spin_2017} for the
density $n_{\rm Yb}$ of Yb atoms in the metastable state,
\begin{equation}
	\dot{n}_{\rm Yb}({\bf r}, t) = - \alpha\, n_{\rm Yb}({\bf r}, t) - \beta\,
	\xi\, n_{\rm Li}({\bf r})\, n_{\rm Yb}({\bf r}, t)\, .
	\label{eqn:DGL}
\end{equation}
Here $\alpha$ is the one-body and $\beta$ the Yb($\triplet$)-Li inelastic loss
rate. The slight reduction of the Li density $n_{\rm Li}$ at the Yb lattice
sites due to the optical lattice potential is described by the density
correction factor $\xi$. It is evaluated to $\xi = 0.65 \pm 0.03$ by the
overlap integral of the Li Bloch wave function and the Yb Wannier
state~\cite{konishi_collisional_2016}. The preparation of localized
Yb($\triplet$) atoms in $n = 1$ sites and the removal of remaining
Yb($\singlet$) atoms ensures that collisions with other Yb atoms are
negligible. Also, by restricting the number of metastable Yb atoms to about
$10\%$ of the total number of Li atoms it is justified to assume $n_{\rm Li}$
as constant during the holding times of interest here. Similarly, the
experiment showed no significant change in the temperature of the Li sample.
The one-body loss rate $\alpha$ is determined to be $\alpha^{-1} = (850
\pm 300)~\msec$ by our independent reference measurements in which Li was
removed from the sample. As discussed in~\cite{schafer_spin_2017} the
inelastic loss rate $\beta$ can be reliably deduced from the initial
exponential decay of the total number of Yb($\triplet$) atoms $N_{\rm Yb}$.
Assuming for short interaction times
\begin{equation}
	N_{\rm Yb}(t) = N_{\rm Yb}(0)\, \exp(-t/\tau_{\rm expt})
	\label{eqn:Nexpt}
\end{equation}
gives rise to an experimental lifetime $\tau_{\rm expt}$. Comparison of the
slope at $t = 0$ of Eq.~(\ref{eqn:Nexpt}) to the initial decay described by
the spatial integral of Eq.~(\ref{eqn:DGL}),
\begin{equation}
	\dot{N}_{\rm Yb}(t = 0) = - \alpha\, N_{\rm Yb}(t=0) - \beta\, \xi\, \int
	n_{\rm Li}({\bf r})\, n_{\rm Yb}({\bf r}, 0)\, {\rm d}^3r\, ,
	\label{eqn:initialDecay}
\end{equation}
yields together with the overlap integral $X = \int n_{\rm Li}({\bf r})\,
n_{\rm Yb}({\bf r}, 0)\, {\rm d}^3r$ the inelastic loss rate
\begin{equation}
	\beta = \frac{N_{\rm Yb}(0)}{\xi\, X} \left( \frac{1}{\tau_{\rm expt}} -
	\alpha \right)\, .
	\label{eqn:beta}
\end{equation}
During the evaluation of $X$ care is taken to include remnant
center-of-mass offsets between the Yb and Li clouds. In the analysis
typically the first seven data points, corresponding to holding times up to
between $2$ and $5~\msec$, are included for the determination of $\tau_{\rm
expt}$; cf.\ dashed lines in Fig.~\ref{fig:decay}.

\begin{figure}[tb!]
	\centering
	\includegraphics[width=8cm]{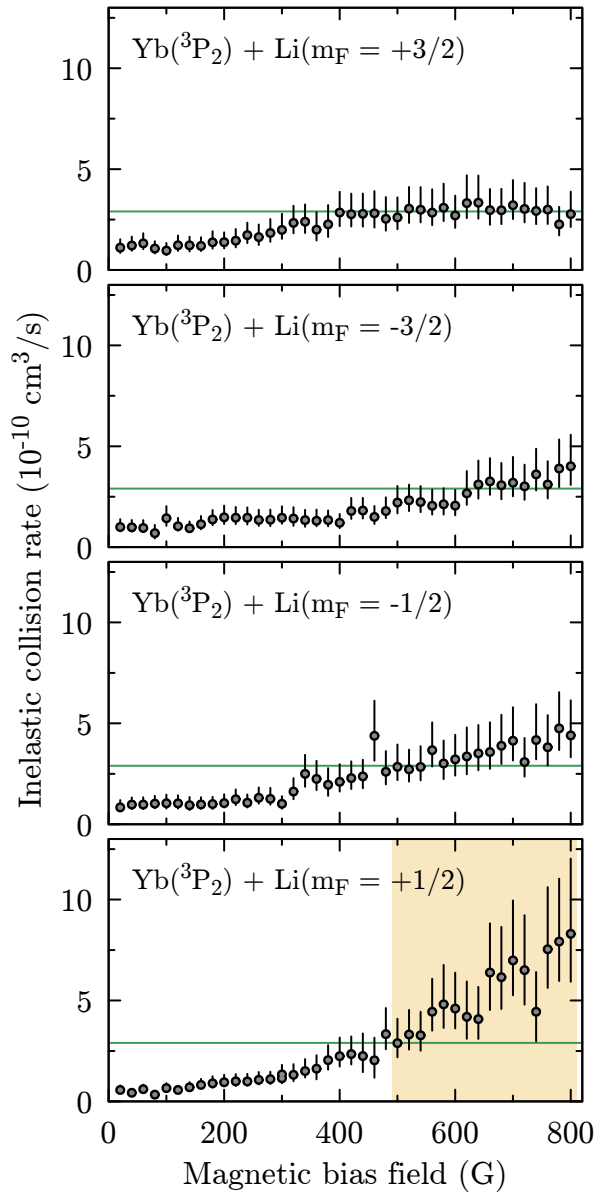}
	\caption{
		Survey of inelastic collision rates for Yb($\triplet, m_J = 0$)-Li($\Li$)
		collisions in magnetic bias field ranges between $20$ and $800~\G$. The Li
		sample is prepared in either the $F = 1/2, m_F = \pm 1/2$
		state (lower two panels) or the $F = 3/2, m_F = \pm 3/2$ state (upper two
		panels). All panels are reproduced in the same scale and a smooth increase
		of the inelastic collision rates is observed in all four cases. No
		distinct resonance like features are found that might indicate the
		presence of an underlying Feshbach resonance. See the main text for a
		discussion of the error bars. Also shown is the expected universal rate
		(green lines) and the shaded area (yellow) marks the parameter range in
		which additionally spectroscopic data on elastic properties was obtained;
		see Sec.~\ref{sec:el}.
	}
	\label{fig:results}
\end{figure}

In a two-step bootstrap approach the uncertainties in the analysis are
evaluated. First, by random resampling and reanalysis of the experimental data
a distribution for $\tau_{\rm expt}$ is determined. Then, repeated evaluation
of Eq.~(\ref{eqn:beta}) while randomly drawing values from the distribution of
$\tau_{\rm expt}$ and from assumed ranges of the remaining parameters a
distribution for $\beta$ is obtained. See~\cite{schafer_spin_2017} for a
detailed discussion and the value ranges used also here. The quantiles at
$50\%$, $15.9\%$ and $84.1\%$ then give the best estimator and a 1-$\sigma$
confidence interval.

The results for the different Li Zeeman states are shown in
Fig.~\ref{fig:results}. At the lowest magnetic field, $20~\G$, we observe
similarly about $1 \times 10^{-10}~\cm^3/\s$ in all four cases. This is in
good agreement with our earlier results~\cite{schafer_spin_2017} obtained at
$0.2~\G$. For increased magnetic fields up to approximately $400~\G$ only a
moderate increase in the inelastic collision rate is observed. Then, up to the
highest fields at $800~\G$, in collisions with Li($F=1/2, m_F=+1/2$) strongly
enhanced losses up to $8^{+4}_{-3} \times 10^{-10}~\cm^3/\s$ are found. The
same tendency can be seen in the remaining cases, albeit to lesser extents,
the inelastic rates staying below $5 \times 10^{-10}~\cm^3/\s$. Pronounced
peaks in the losses, indicative of possible Feshbach resonances, are not
observed.
\begin{figure}[H]
	\centering
	\includegraphics[width=8cm]{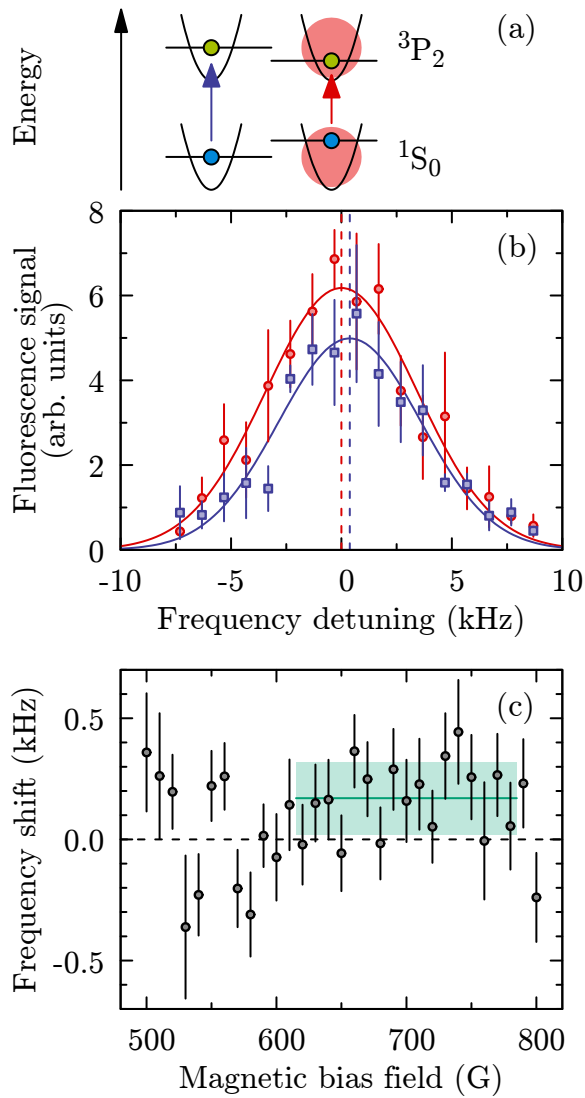}
	\caption{
		Direct spectroscopy of Yb($\triplet$)-Li($F=1/2, m_F=+1/2$) interspecies
		interaction. 
		(a) Sketch of the measurement principle where a change in the interaction
		energy between Li (red) and Yb in either $\singlet$ (blue) or $\triplet$
		(green) state leads to a shift of the transition energy (arrows).
		(b) Excitation spectra for singly occupied lattice sites at a
		lattice depth of $15\,E_R^{\rm Yb}$ and at $500~\G$ magnetic bias field.
		With respect to the reference measurement without Li (red circles) a
		reduction of the peak height and a slight shift towards higher excitation
		frequencies in the case with Li (blue squares) is observed. The data are
		fitted by Gaussian functions (lines). Error bars denote the standard error
		from five measurements and the frequency detuning is set to zero when in
		resonance with the reference measurements.
		(c) Summary of the spectroscopic line shifts obtained in the
		range $500$ to $800~\G$. Data were taken in steps of $10~\G$ and the
		frequency shifts with respect to a corresponding reference measurement
		without Li obtained by Gaussian fitting are reported here together with
		the fitting 1-$\sigma$ error estimates. The dashed line is to mark
		vanishing frequency shifts. In the range $620$ to $780~\G$ a small
		positive frequency shift is observed (green line and shaded areas as
		1-$\sigma$ confidence interval).
	}
	\label{fig:spectroscopy}
\end{figure}

\subsection{Elastic interspecies interactions}
\label{sec:el}

Using interspecies thermalization measurements~\cite{ivanov_sympathetic_2011,
hara_quantum_2011} and an analysis of the bosonic dipole oscillation frequency
shift in a superfluid Yb-Li mixture~\cite{roy_two-element_2017} the Yb-Li
ground-state scattering length is known to be $a_{{\rm Yb}(\singlet){\rm Li}}
= (+15 \pm 2)\,a_0$, where $a_0$ is the Bohr radius. In contrast, the present
work probes the interspecies scattering length $a_{{\rm Yb}(\triplet){\rm
Li}}$ between metastable Yb and Li. By measuring the frequency spectrum of the
Yb($\triplet$) Mott-insulator $n=1$ shell direct access to on-site interaction
energies is gained. By comparison of the resonance positions in the presence
and absence of Li a sensitive probe to the interspecies
interaction~\cite{bloch_many-body_2008},
\begin{equation}
	U_{\rm YbLi} = \frac{2\pi \hbar^2\, a_{\rm YbLi}}{m_{\rm red}} 
	\int |w_{\rm Yb}({\bf r})|^2 |\psi_{\rm Li}({\bf r})|^2 n_{\rm Li}({\bf r})\,
	{\rm d}^3{\bf r}\ ,
	\label{eqn:Ufull}
\end{equation}
is obtained. Here, $a_{\rm YbLi}$ is the interspecies scattering length,
$m_{\rm red}$ is the reduced mass, and $w_{\rm Yb}$ and $\psi_{\rm Li}$ are
the Yb Wannier and Li Bloch wave functions. The integral is taken over the
volume of a single lattice site and simplifies to
\begin{equation}
	U_{\rm YbLi} = \frac{2\pi \hbar^2\, a_{\rm YbLi}}{m_{\rm red}}\, \xi\,
	n_{\rm Li}\, .
	\label{eqn:Usimpl}
\end{equation}
Our experimental signal is sensitive to a shift of the spectral line and thus
to a change in the interaction energy $U_{{\rm Yb}(\triplet){\rm Li}} -
U_{{\rm Yb}(\singlet){\rm Li}}$ as depicted in Fig.~\ref{fig:spectroscopy}(a).
Assuming typically $n_{\rm Li} = 3 \times 10^{12}~\cm^{-3}$ one finds that a
resonance shift of $+0.5~\kHz$ corresponds to a change in the scattering
length by $a_{{\rm Yb}(\triplet){\rm Li}} - a_{{\rm Yb}(\singlet){\rm Li}}
\approx 23~\nm = 440\,a_0 $.

Data was taken for magnetic fields $> 500~\G$ where magnetic mixing with the
$^3{\rm P}_{1}$ state ensured a strong spectroscopic signal. We concentrated
on the Yb($\triplet, m_J = 0$)-Li($\singlet, F=1/2, m_F=+1/2$) case where the
largest change of the inelastic rate for high fields was observed, see shaded
region in Fig.~\ref{fig:results}. The spectroscopic data at $500~\G$ is shown
in Fig.~\ref{fig:spectroscopy}(b). The fluorescence signal with Li (blue)
generally shows a lower amplitude than the data without it (red). This is due
to the partial decay of the Yb($\triplet$) atoms during excitation
($0.3~\msec$), removal of Yb($\singlet$) atoms ($0.3~\msec$) and repumping
($1~\msec$). Of importance here, however, is a possible shift of the
spectroscopic line. Gaussian fits to the data reveal in this case a frequency
up-shift by $(0.4 \pm 0.2)~\kHz$, corresponding to $a_{{\rm Yb}(\triplet){\rm
Li}} - a_{{\rm Yb}(\singlet){\rm Li}} = (350 \pm 180)~a_0$. In
Fig.~\ref{fig:spectroscopy}(c) we summarize the measured frequency shifts up
to $800~\G$. Even though the inelastic rate shows variations by a factor of
two in this range the observed frequency shifts remain below $0.5~\kHz$
limiting the change of $a_{\rm YbLi}$ to below $400\,a_0$. Between $620$ and
$780~\G$ a positive frequency shift of $(0.17 \pm 0.15)~\kHz$ is observed
(cf.\ shaded area in Fig.~\ref{fig:spectroscopy}) which corresponds to an
increase of the interspecies scattering length by $(150 \pm 130)\,a_0$.

\section{Discussion and Conclusion}
\label{sec:discussion}

Using high-resolution spectroscopic measurements we analyze both the inelastic
collisional rates and the elastic interspecies interactions between
Yb($\triplet, m_J = 0$) atoms and four Zeeman states of ground state Li for
magnetic field strengths up to $800~\G$. We find in general moderate inelastic
loss rates of about $10^{-10}~\cm^3/\s$ at low fields. At our highest fields
increased rates are found that still remain below $10^{-9}~\cm^3/\s$. The
increase is generally found to be smooth and picks up strength for fields
beyond about $400~\G$. No resonance like structures are observed. Focusing on
a parameter range with a particular strong change in the inelastic properties
we directly investigate the on-site Yb($\triplet, m_J=0$)-Li($F=1/2,
m_F=+1/2$) interaction energy. The observed frequency shifts remain below $\pm
0.4~\kHz$. In the theoretical work
Ref.~\cite{gonzalez-martinez_magnetically_2013} generally, even at predicted
strong Feshbach resonances, $|a_{{\rm Yb}(\triplet){\rm Li}}| < 1000\,a_0$ is
found corresponding to a resonance shift of about $1~\kHz$. At other magnetic
fields scattering lengths remaining below $200\,a_0$ are predicted. In the
range between $620$ and $780~\G$ a systematic, non-zero shift prevails that
indicates a small but finite increase of $a_{{\rm Yb}(\triplet){\rm Li}}$ with
respect to $a_{{\rm Yb}(\singlet){\rm Li}}$ by about $150\,a_0$.

The inelastic collision rates reported here are generally higher than those
theoretically predicted in~\cite{gonzalez-martinez_magnetically_2013} for
collisions with Yb($\triplet, m_J=-2$) and go beyond the estimated universal
rate~\cite{idziaszek_universal_2010, gonzalez-martinez_magnetically_2013} of
$2.9 \times 10^{-10}~\cm^3/\s$; cf.\ green lines in Fig.~\ref{fig:results}.
The observed strong losses highlight the effectiveness of relaxation processes
driven by the anisotropic Yb($\triplet$)-Li interaction potential. The
measured on-site Yb($\triplet$)-Li interaction energy confirms the expected
moderate variability~\cite{gonzalez-martinez_magnetically_2013} of $a_{{\rm
Yb}(\triplet){\rm Li}}$. The absence of resonant-like behavior in all our data
is striking and should motivate corresponding studies. The survey of
parameters presented in the current paper is to serve as a basis for a
discussion on the coarse structure of the Yb($\triplet$)-Li interactions. In
the future, experiments with greater resolution might reveal resonance
structures with small widths that were beyond the scope of the present
research. It will be further theoretically of great interest and
experimentally challenging to expand the methods presented here to
measurements of the energetically lowest but strongly magnetic-field-dependent
$m_J = -2$ substate of Yb($\triplet$).

\section*{Acknowledgments}
The authors would like to thank A.\ Kell for invaluable help in an early
stage of the experiment. This work was supported by the Grant-in-Aid for
Scientific Research of JSPS Grants No.\ JP25220711, No.\ JP26247064, No.\
JP16H00990, No.\ JP16H01053, and No.\ JP17H06138, JST CREST Grant No.\
JPMJCR1673 and the Impulsing Paradigm Change through Disruptive Technologies
(ImPACT) program by the Cabinet Office, Government of Japan. H.\ K.\
acknowledges support from JSPS.

\end{document}